\documentclass[apj,numberedappendix,epsfigure]{emulateapj}

\shorttitle{PULSATIONS ON HE CORE WDS}
\shortauthors{STEINFADT, BILDSTEN, \& ARRAS}

\newcommand{\be}{\begin{eqnarray}}
\newcommand{\ee}{\end{eqnarray}}

\begin{document}


\title{Pulsations in Hydrogen Burning Low Mass Helium White Dwarfs}

\slugcomment{The Astrophysical Journal, 718:441 - 445, 2010 July 20}

\author{Justin D. R. Steinfadt}
\affil{Department of Physics, Broida Hall,\\University of California, Santa Barbara, CA 93106, USA;\\jdrs@physics.ucsb.edu}

\author{Lars Bildsten}
\affil{Kavli Institute for Theoretical Physics and Department of Physics, Kohn Hall,\\University of California, Santa Barbara, CA 93106, USA;\\bildsten@kitp.ucsb.edu}

\author{Phil Arras}
\affil{Department of Astronomy,\\University of Virginia, P. O. Box 400325, Charlottesville, VA 22904, USA;\\arras@virginia.edu}






\begin{abstract}

Helium core white dwarfs (WDs) with mass $M\lesssim 0.20\,M_{\odot}$ undergo several Gyrs of stable hydrogen burning as they evolve.  We show that in a certain range of WD and hydrogen envelope masses, these WDs may exhibit $g$-mode pulsations similar to their passively cooling, more massive carbon/oxygen core counterparts, the ZZ~Cetis.  Our models with stably burning hydrogen envelopes on helium cores yield $g$-mode periods and period spacings longer than the canonical ZZ~Cetis by nearly a factor of 2.  We show that core composition and structure can be probed using seismology since the $g$-mode eigenfunctions predominantly reside in the helium core.  Though we have not carried out a fully nonadiabatic stability analysis, the scaling of the thermal time in the convective zone with surface gravity highlights several low-mass helium WDs that should be observed in search of pulsations:  NLTT~11748, SDSS~J0822+2753, and the companion to PSR~J1012+5307. Seismological studies of these He core WDs may prove especially fruitful, as their luminosity is related (via stable hydrogen burning) to the hydrogen envelope mass, which eliminates one model parameter.

\end{abstract}



\section{Introduction}

White dwarfs (WDs) are observed to pulsate in normal modes of oscillation ($g$-modes) which are determined by the structure of the stellar interior and atmosphere \citep{win08}. Those with hydrogen atmospheres exhibit pulsations when they enter the ZZ Ceti variable (DAV) instability region, a discrete strip in the $T_{\rm eff}$--$\log g$ plane that spans $11,000\,$K $ \lesssim T_{\rm eff} \lesssim 12,250$\,K at $\log g \approx 8.0$.  The ZZ~Ceti strip has been investigated both theoretically \citep{bra97,wu99,fon03} and empirically \citep{wes91,muk04,cas07,gia07}.  To date, all known ZZ~Ceti pulsators have masses $>0.5\,M_\odot$, implying cores composed of carbon, oxygen, and heavier elements.

Lower mass ($M<0.5\,M_\odot$) WDs with nearly pure helium cores are made on the red giant branch (RGB) when core growth is truncated before reaching $\approx$0.45$-$0.47$M_{\odot}$ ($\log g \approx 7.67$ at $T_{\rm eff} \approx 11,500$\,K),  prior to the helium core flash  \citep{dcr96,dom99,pie04}.  Two modes of envelope mass loss can cause this:  strong  winds or binary interaction.  Significant mass loss due to stellar winds in high metallicity systems may strip the H envelope,  preventing the He core flash \citep{dcr96,han05}.  Common envelopes induced by binary interactions also lead to significant mass loss \citep{ibe93,mar95}, and  make very low-mass He WDs ($M<0.2\,M_{\odot}$) when the binary interaction occurs at the base of the RGB \citep{van96,cal98,bas06}.  Thus, He is the expected core composition for WDs below $\approx$0.45$-$0.47$\,M_{\odot}$. However, very little direct evidence exists of He cores. The overbrightness of old WDs \citep{han05} in the star cluster \object[NGC 6791]{NGC~6791} \citep{bed05} presents possible evidence.  The detection of low $\log g$ young WDs make a plausible argument for the old WDs to be He core \citep{kal07}, however, other possible explanations remain \citep{del02,bed08a,bed08b,gar10}.

Asteroseismology offers the possibility of directly constraining the He core composition in these  low-mass WDs, as the $g$-mode periods  provide information on WD mass, mass of H envelope, and core
composition \citep{cor02,cas08}. Theoretically, these principles have been applied to C/O versus O/Ne core WDs by \citet{cor04} and O versus He core WDs by \citet{alt04}.  However, to carry this out, we need
to find pulsating He core WDs, something that has yet to occur.

Past studies have illuminated a dichotomy in the evolution of the He core WDs \citep{dri99,ser02,pan07} that impacts their seismic properties and prevalence as pulsators.  For masses $\gtrsim$0.2\,$M_{\odot}$ (dependent upon metallicity) the H envelope experiences a multitude of H shell flashes that reduces its mass, eventually allowing the WD to cool rapidly. Such objects traverse the extrapolated ZZ~Ceti instability strip in $\sim10$--$100$ Myr \citep{pan07}, allowing for an investigation of their H layer mass, and confirmation of pure helium core. However, there are presently no known WDs in, or near, the extrapolated strip for masses in the $0.2\,M_{\odot}<M<0.5\,M_{\odot}$ range \citep{ste08}, inhibiting such research. Less massive ($<0.2\,M_\odot$) helium WDs have a different evolution, undergoing stable H burning for Gyr, slowing their evolution to rates that may yield more in the extrapolated instability strip. However, the presence of a thick, actively burning hydrogen layer requires new seismic modeling, especially for the eventual assessment of the of the instability strip for these unusual WDs.  The recent discovery of three such objects \citep{kaw09,kil09} makes our work quite timely.

Motivated by a desire to study the pulsational properties of these long-lived systems, we begin in Section\ref{sec:bmod} by constructing He core WD models with stable H burning shells flexible enough for seismic investigations, and compare to the results from evolutionary codes \citep{ser02, pan07}. We discuss the unusual seismic properties of these objects in Section \ref{sec:panal}, where we calculate their adiabatic mode structure and, using an approximation for the instability criterion of Brickhill's theory \citep{bri91}, highlight the potential  location of the He core WD instability strip. In Section \ref{sec:conc}, we suggest a few intriguing pulsation candidates amongst the very lowest mass WDs \citep{van96,cal98,bas06,kaw06,kaw09,kil09}, where our simple models apply. We close by highlighting the need for future work, especially if observations of our suggested targets yield the first pulsating, low-mass, He WD.  


\section{Hydrogen Burning Models}
\label{sec:bmod}

\begin{figure}[t]
	\centering
	\epsscale{1.0}
	\plotone{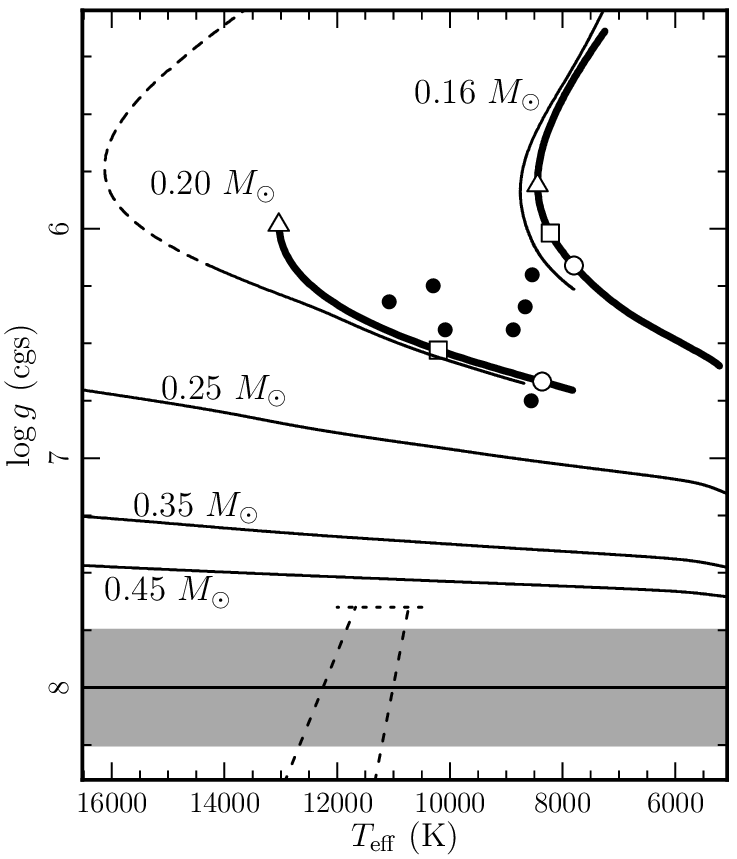}
	\caption{Comparison of our model tracks, the heavy black lines, against those of \citet{pan07}, $0.16\,M_{\odot}$, and \citet{ser02}, 0.20, 0.25, 0.35, and 0.45$\,M_{\odot}$, the thin black lines.  Our tracks deviate significantly at low gravity from those of \citet{ser02}.  The solid circles are several candidate $<0.20\,M_{\odot}$ WD pulsation candidates \citep{van96,cal98,bas06,kaw06,kil09,kaw09}.  Note their location on the higher gravity, past the $T_{\rm eff}$ turn-around portion of the models.  The dashed lines in the \citet{ser02} models denote where the CNO luminosity is greater than 10\% of the PP~chains.  This explains the deviation as our models only have PP-chain luminosity, appropriate to the most relevant regions at higher gravity.  Five Gyr have elapsed between the triangle and square marks and the square and circle marks.  The gray area approximates the location of the C/O WDs with the vertical dashed lines denoting the empirical ZZ~Ceti instability strip \citep{gia07}.  The dotted line above the empirical instability strip denotes the gravity of the lowest known gravity ZZ~Ceti, HE~0031-5525 \citep{cas06}. }
	\label{fig:compare}
\end{figure}

The $<0.2\,M_{\odot}$ He core WDs of interest for our work are undergoing stable H burning via the PP~chains in a low-mass shell.  The solid/dashed lines in Figure \ref{fig:compare} are the models from \citet{ser02} and \citet{pan07}.  These lines transition to solid when 90\% of the luminosity is generated from the PP~chains.  The solid circles are the locations of the observed WDs of interest here, and are clearly in a region dominated by PP-chain burning.  For this reason, we construct models with only PP-chain burning.  In addition, at these late times, the diffusive timescale at the burning zone is much shorter than the age and the WD core temperature is set by that in the stable burning layer.  For  these reasons, the prior evolution of a stably burning WD does not affect its properties at this stage of evolution.

We construct models of a stably burning H envelope on an He core, by solving the equations of hydrostatic balance, heat transport, energy generation, and mass conservation.  Between the H and He layers the most important physics in our models is the chemical profile.  In these layers diffusive equilibrium is valid as the evolutionary timescale (dominated by nuclear burning, $\sim1$--$10$\,Gyr) is significantly longer than the diffusive timescale ($\sim10$--$100$\,Myr) over a pressure scale height.  We derive the equilibrium electric field by assuming each species is in hydrostatic balance with gravity and the electric force and charge neutrality (see \citet{cha03} for more detailed derivation).  With this electric field we generate an additional differential equation for one of the chemical species (charge neutrality gives the rest) to be simultaneously solved with the equations of stellar structure.  Given our set of differential equations and boundary conditions, our model reduces to two parameters, total mass and total H mass, fewer than those for passively cooling C/O WDs which require total mass, H mass, He mass and surface temperature.

Our models contain zero metallicity.  Additional elements require additional differential equations for their diffusive profile for which equilibrium conditions may not exist.  Therefore, the PP-chains ($p(p,e^+\nu_e)^2{\rm H}(p,\gamma)^3{\rm He}(^3{\rm He},pp)^4{\rm He}$ or $^3{\rm He}(^4{\rm He},\gamma)^7{\rm Be}(e^-,\nu_e(\gamma))^7{\rm Li}(p,\alpha)^4{\rm He}$ or $^7{\rm Be}(p,\gamma)^8{\rm B}(e^+\nu_e)^8{\rm Be}\;2\;^4{\rm He}$) are the only source of nuclear energy.  We assume $^3$He has reached its equilibrium abundance, peaking at $\sim 10^{-4}$ to $10^{-3}$ by mass.  We generate evolutionary tracks by conserving total mass; nuclear burning converts envelope mass into core mass.  Figure \ref{fig:compare} compares our models with the time-dependent models of \citet{ser02} and \citet{pan07}, exhibiting discrepancy at low gravity (large envelope mass) but excellent agreement at high gravity (low envelope mass).  This is attributed to large CNO luminosities at high envelope mass in the non-zero metallicity models of \citet{ser02}.  At low envelope masses, the core temperatures have lowered to $\approx10^{7}$\,K and CNO elements have diffused out of the burning region \citep{pan07}.  Therefore, PP-chain luminosity dominates so we expect our models to be valid in this regime, where, as we show, the candidate He core pulsating WDs are likely to be found.  As nuclear burning determines the evolution of our models, the timescales are long, of order several Gyr.  Figure \ref{fig:compare} illustrates the evolution of our models over 5 (triangles to squares) and 10 Gyr (triangles to circles) from the point of maximum $T_{\rm eff}$.

The microphysics, opacities, equation of state, and nuclear energy generation, are all handled by the Modules for Experiments in Stellar Astrophysics (MESA)\footnote{http://mesa.sourceforge.net} code, developed by B. Paxton et al. (2010, in preparation).  Within MESA, the opacities are drawn from OPAL \citep{ing93,ing96}, the \citet{fer05} low-temperature tables, and the \citet{cass07} electron conduction tables.  The equation of state is derived from OPAL \citep{rog02}, low-temperature SCVH \citep{sau95}, and fully ionized high temperature and density HELM \citep{tim00}.  Nuclear energy generation is calculated using the techniques developed by \citet{tim99}.


\section{Non-Radial Pulsation Analysis}
\label{sec:panal}

To analyze the non-radial pulsational properties of our He-core WD models we perturb and linearize the fluid equations of momentum, energy, and mass conservation.  We set the transverse wavenumber of order $\ell$ as $k_h^2 = \ell ( \ell + 1 ) / r^2$.

\subsection{WKB Approximation}

The star is divided into regions of wave propagation and evanescence. In the propagation zone, the wavelength $k_r^{-1}$ is much smaller than the characteristic length scales associated with the background, such as the radius $r$, near the center, and the pressure scale height $\lambda_P=p/(\rho g)$ near the surface.  This allows for the WKB approximation where all state variables are $\propto \exp(i \int^{r'} dr k_r)$.  Neglecting perturbations on the gravitational field (the Cowling approximation) we further reduce the linearized pulsation equations into the dispersion relation,
\be
	k_r^2 = \frac{\left( N^2 - \omega^2 \right) \left( c_s^2 k_h^2 - \omega^2 \right)}{\omega^2 c_s^2}, \label{eqn:disprel}
\ee
\noindent where $\omega$ is the frequency of pulsation, $N$ is the Brunt--V\"ais\"al\"a frequency,
and $c_s$ is the adiabatic sound speed.  For propagating waves, Equation (\ref{eqn:disprel})
defines the resonant cavity for waves of two types.  When $\omega > N$ and $c_s k_h$ (the Lamb
frequency), waves propagate as sound waves ($p$-modes) where pressure provides the restoring force.
When $\omega < N$ and $c_s k_h$, waves propagate as gravity waves ($g$-modes) where gravity
provides the restoring force.  \citet{bri83} showed that convective driving could drive the
amplitude of $g$-mode pulsations to an observable level; these are the pulsations observed
in the ZZ Cetis \citep{war72}.

\begin{figure}[t]
	\centering
	\epsscale{1.0}
	\plotone{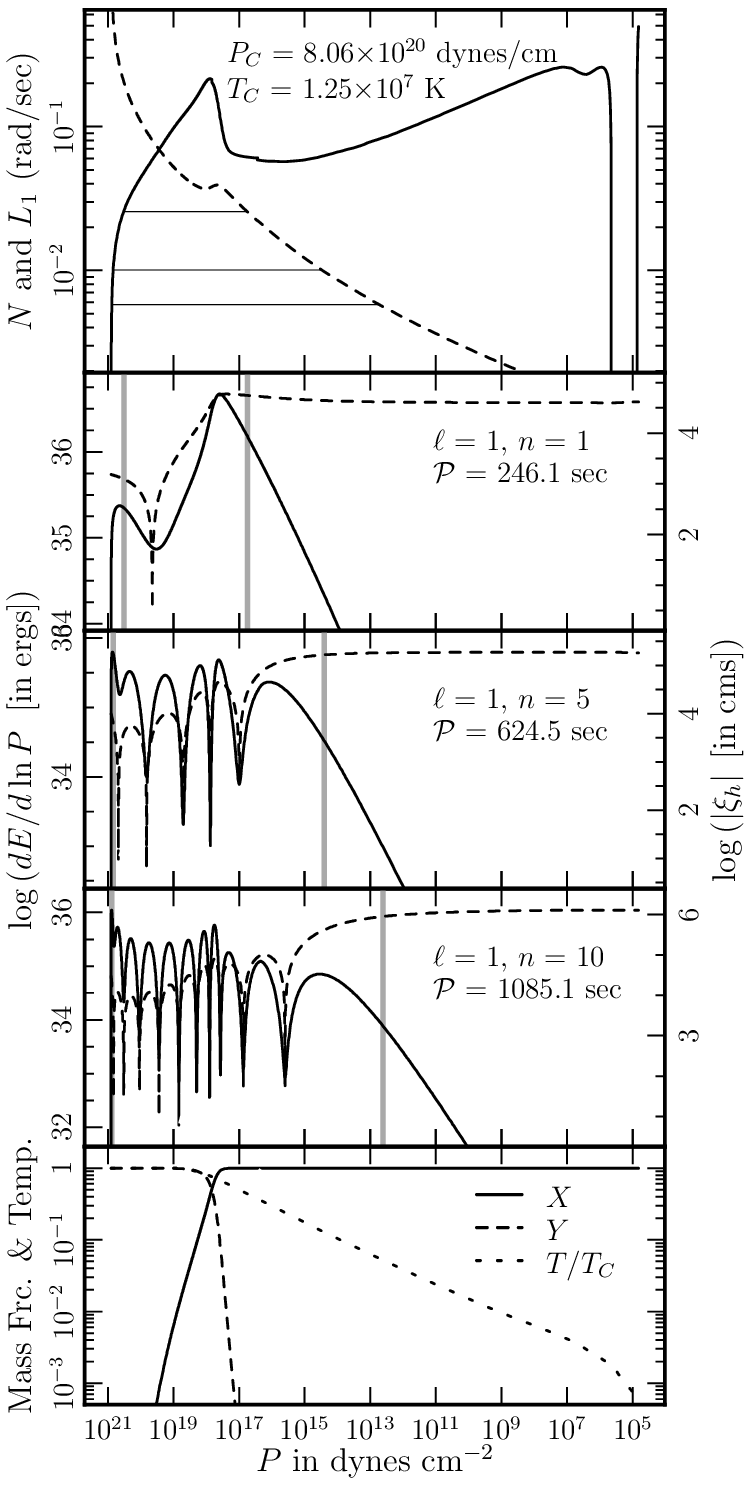}
	\caption{Propagation diagram (upper panel) for our $M_{\rm WD} = 0.17\,M_{\odot}$ and $M_{\rm env} = 3.15 \times 10^{-3}\,M_{\odot}$ model.  Gravity mode ($g$-mode) pulsations exist below both the Brunt--V\"ais\"al\"a frequency (solid curve) and the Lamb frequency ($\ell = 1$; dashed curve).  The thin horizontal lines delineate the locations of the $n = 1$, 5, and 10 modes for $\ell = 1$.  Note the large bump in the Brunt--V\"ais\"al\"a frequency due to the Ledoux contribution (Equation \ref{eqn:led}) that aligns with the composition transition region (bottom panel).  The middle three panels show the eigenfunction solutions of $\xi_h$ (dashed curves) and mode energy (solid curves) for the $n = 1$, 5, and 10 modes for $\ell = 1$.  The gray vertical lines denote the boundaries of the WKB propagation cavity.  These show the bulk of each mode to reside in the core, below the H/He transition, with only a couple nodes existing in the envelope.  This model is our closest fit to the properties of NLTT~11748 \citep{kaw09}.}
	\label{fig:propdiag}
\end{figure}

We approximate the frequencies of propagating $g$-modes using the WKB quantization condition $\int_{r_{\rm in}}^{r_{\rm out}}dr k_r = n \pi$. Under the assumptions $\omega \ll N$ and $\omega \ll c_s k_h$, Equation (\ref{eqn:disprel}) gives,
\be
	\omega_{n, \; \ell} = \frac{ \sqrt{ \ell (\ell+1) } }{n \pi} \int_{r_{\rm in}}^{r_{\rm out}} \frac{dr}{r} N, \label{eqn:omega}
\ee
\noindent where the integral is bounded by the frequency dependent resonant cavity. Here 
$r_{\rm in}$ and $r_{\rm out}$ are the radii where $\omega=N$ and $\omega=c_s k_h$, respectively; see Figure \ref{fig:propdiag} for illustration.  Under these assumptions, the derived mode periods are only accurate for large radial order, $n \gg 1$.  Figure \ref{fig:propdiag} shows a propagation diagram for an $M_{\rm WD} = 0.17\,M_{\odot}$ and $M_{\rm env} = 3.15\times10^{-3}\,M_{\odot}$ model.  As is evident, the resonant cavity for the higher order $g$-modes samples much of the core and envelope while the lower orders are most affected by the transition region.  It is obvious that the contribution of the transition region is quite important, therefore, close attention must be paid to the Brunt--V\"ais\"al\"a frequency.  For changing composition, the Brunt--V\"ais\"al\"a frequency is,
\be
	N^2 = \frac{g}{\lambda_P} \left[ \frac{\chi_T}{\chi_{\rho}} \left( \nabla_{\rm ad} - \nabla \right) + B \right],
\ee
\noindent where, $\chi_{\rho} \equiv \left. \partial \ln P / \partial \ln \rho \right|_{T, \; \left\{ X_i \right\} }, \chi_{T} \equiv \left. \partial \ln P / \partial \ln T \right|_{\rho, \; \left\{ X_i \right\} }$, and,
\be
	B = \sum_{i=1}^{I-1} \left. \frac{\partial \ln \rho}{\partial X_i} \right|_{T, \; P, \; \left\{ X_{j \neq i} \right\}} \frac{d X_i}{d \ln P}, \label{eqn:led}
\ee
\noindent is the compositionally dependent Ledoux term (modified from \citealp{bra91}) which accounts for the bulk of the bump in the Brunt--V\"ais\"al\"a frequency at the composition transition zone in Figure \ref{fig:propdiag}.

\subsection{Numerical Analysis}

\begin{figure*}[t]
	\centering
	\epsscale{1.0}
	\plotone{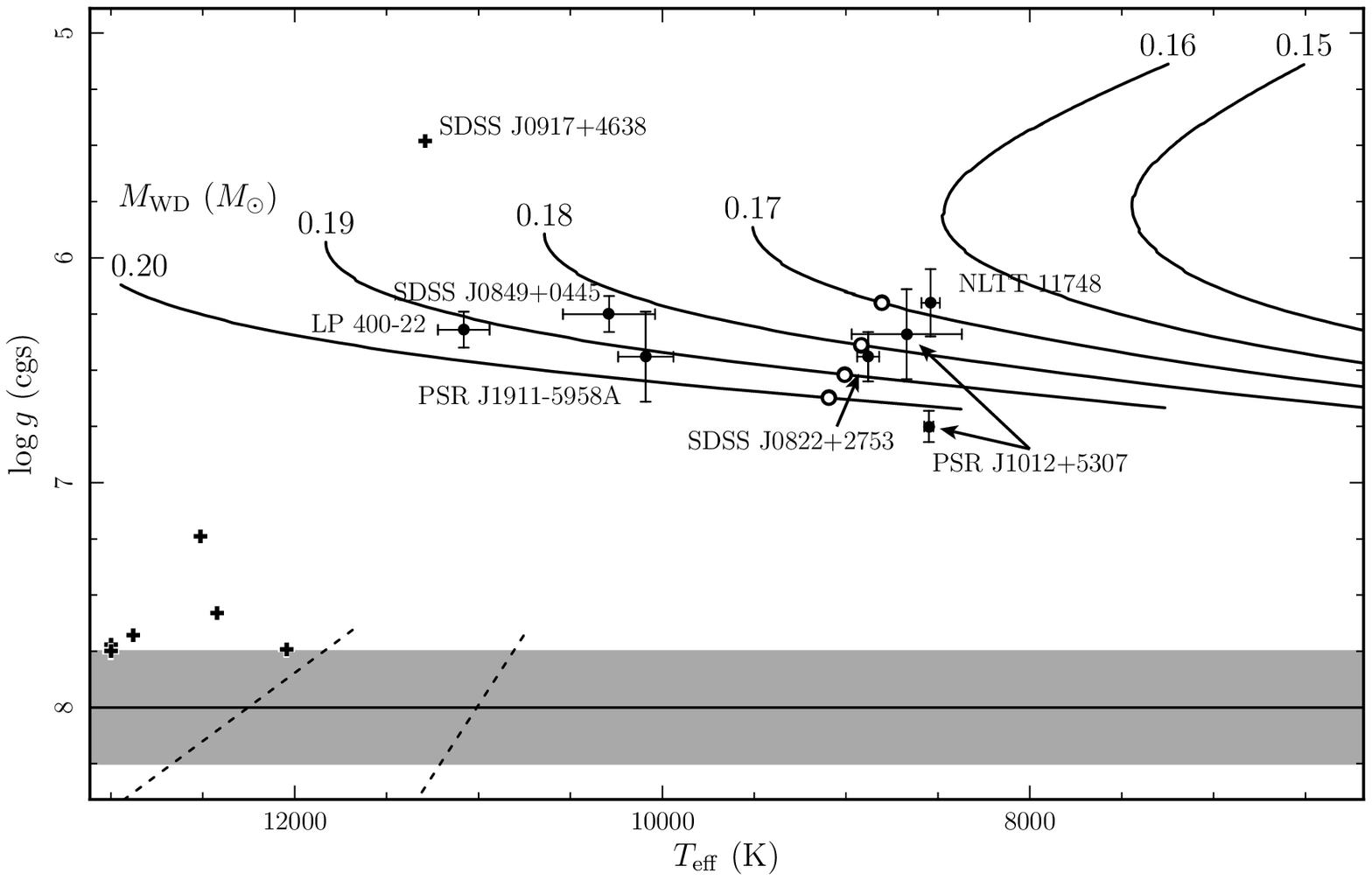}
	\caption{Evolution of stably burning low-mass He cores.  Evolution proceeds top to bottom.  We start our $0.20\,M_{\odot}$ model where the deviations from the time-dependent \citet{ser02} models have been reduced to less than $\Delta \log g = 0.2$ (see Figure \ref{fig:compare}).  For our 0.17, 0.18, and $0.19\,M_{\odot}$ models we begin at the maximum $T_{\rm eff}$ turn as we have no models to directly compare.  We show our entire 0.15 and $0.16\,M_{\odot}$ models, starting with $M_{\rm env} = 10^{-2}\,M_{\odot}$, as our comparison with \citet{pan07} show them to be reliable (Figure \ref{fig:compare}).  For comparison we exhibit the location of the C/O WDs (the shaded region) and their associated empirical instability strip, the sloped dashed lines \citep{gia07}.  Candidate objects are drawn from several sources. The bullets are PSR~J1911-5958A \citep{bas06}, PSR~J1012+5307 \citep{van96,cal98}, SDSS~J0849+0445 and SDSS~J0822+2753 \citep{kil09}, NLTT~11748 \citep{kaw09}, and LP~400-22 \citep{kaw06}.  The crosses are those SDSS selected low-mass WDs confirmed by MMT spectra from \citet{kil07}.  The circles on the evolution curves represent where the convective thermal time instability criterion is met for the $\ell=1$ and $n=1$ modes.}
	\label{fig:inststrip}
\end{figure*}

To obtain more accurate mode periods that are not restricted to the high radial orders as our WKB analysis is, we must turn to the boundary value problem for adiabatic pulsations.  The equations and method are described in \citet{unn89}.  Perturbation of the equations reduce to the following three equations for the pressure perturbation $\delta p \equiv \rho \psi$, the radial Lagrangian displacement $\xi_r$ (the transverse Lagrangian displacement is $\xi_h = \psi r / \omega^2$), and the gravitational potential perturbation $\delta \phi$:
\be
	\frac{d\psi}{dr} &=& (\omega^2 - N^2) \xi_r + \frac{N^2}{g} \psi - \frac{d \delta \phi}{dr}, \label{eqn:dpsidr} \\
	\frac{d\xi_r}{dr} &=&  \left( \frac{g}{c_s^2} - \frac{2}{r} \right) \xi_r + \left(
\frac{k_h^2}{\omega^2} - \frac{1}{c_s^2} \right) \psi +  \frac{k_h^2}{\omega^2} \delta \phi , \label{eqn:dxirdr}
\ee
\be
	\frac{1}{r^2} \frac{d}{dr} \left( r^2 \frac{d \delta \phi}{dr} \right) &=& 4 \pi G \rho \left( \frac{1}{c_s^2} \psi + \frac{N^2}{g} \xi_r \right) + k_h^2 \delta \phi . \label{eqn:ddphidr}
\ee
\noindent At the center of the star, requiring the variables to be finite leads to the scalings $\psi \propto r^\ell$ and $\xi_r \propto r^{\ell-1}$. Placing these scalings into Equations (\ref{eqn:dpsidr}), (\ref{eqn:dxirdr}), and (\ref{eqn:ddphidr}) leads to the central boundary conditions,
\be
	\omega^2 \xi_r &=& \ell \frac{\psi + \delta \phi}{r}, \label{eqn:bc1} \\
	\ell \frac{\delta \phi}{r} &=& \frac{d \delta \phi}{dr}, \label{eqn:bc2}
\ee
\noindent at a nonzero, but small, radius $r$.  At the surface, we require the perturbations to be both finite and upwardly evanescent.  \citet{unn89} show that the correct boundary conditions are then
\be
	\psi &=& g \xi_r, \label{eqn:bc3} \\
	-(\ell + 1) \frac{\delta \phi}{r} &=& \frac{d \delta \phi}{dr}, \label{eqn:bc4}
\ee
\noindent at the upper boundary of the model. We solve Equations (\ref{eqn:dpsidr}), (\ref{eqn:dxirdr}), and (\ref{eqn:ddphidr}) with the boundary conditions in Equations (\ref{eqn:bc1}), (\ref{eqn:bc2}), (\ref{eqn:bc3}), and (\ref{eqn:bc4}) using the shooting method to obtain all mode periods.

The middle three panels in Figure \ref{fig:propdiag} display the transverse displacement eigenfunctions and energy density  for the $n = 1$, 5, and 10 modes for $\ell = 1$.  The energy density illustrates where each mode ``lives,'' that is, what portions of the star most affect the mode period.  These modes live primarily in the core, below the H/He transition region, as the energy density declines rapidly in the lower pressure H layer.  This contrasts the normal ZZ~Cetis \citep{fon08}.  This is predominantly because the Brunt--V\"ais\"al\"a frequency in the He core is larger than in the envelope (see Figure \ref{fig:propdiag}, also noted in \citealt{alt04}). The electron degeneracy in all WD cores leads to most of the entropy in the ions, yielding $N^2 \sim A^{-1} (k_bT/E_F)(g/\lambda_P)$, where $A$ is the ion mass and $E_F$ is the electron Fermi energy. Hence, there are two reasons why $N$ is relatively larger in an He core than in a C/O core. First, low-mass implies smaller $E_F \propto M^{4/3}$, and  second, $1/A$ is larger for He than a C/O mixture.  This shows the power these modes will have in probing the core composition once pulsations have been detected and accurate periods measured.

The observability of these modes requires that they be driven.  One type of driving mechanism requires that a portion of the escaping heat flux be converted into the mechanical energy of the pulsation modes.  An example is given by the $\kappa$-mechanism \citep{dzi77,dzi81,dol81,win82}, in which the rapid outward increase in opacity associated with an ionization zone bottlenecks the heat flux. For ZZ~Ceti-like pulsations, \citet{bri83} proposed that the response of the convection zone itself to the pulsation drives the instability \citep{bri83,bri91,wu99}.  For now we consider only the convective driving mechanism\footnote{The existence of the nuclear burning region inside the mode propagation region may additionally drive modes due to the sensitivity of the nuclear reactions to temperature, the $\epsilon$-mechanism \citep{kaw88}.  We have yet to investigate this possibility.}.  If the convection zone can thermally adjust on a timescale shorter than the pulsation period, $\mathcal{P}$, then the pulsation will be damped.  This motivates the convective thermal time instability criterion $\mathcal{P} \le 8 \pi \tau_{\rm th, \; bcvz}$, where $\tau_{\rm th, \; bcvz}$ is the thermal time from the base of the convection zone to the surface \citep{bri91,wu99}.  \citet{wu99} show (in their Figure 7) this criterion to be quite accurate in the high $T_{\rm eff}$ (the ``blue'' edge) limit compared to fully nonadiabatic calculations on a $\log g = 8.0$ WD.  However, the He core WDs are at significantly lower gravity, requiring an extrapolation in $\log g$.  Calculations show that the rapid increase in $\tau_{\rm th, \: bcvz}$ due to the deepening convective zone occurs at lower $T_{\rm eff}$ for lower $\log g$, nearly $1500$\,K for $\log g = 8$ to $\log g = 7$.  The circle points on the $\log g$--$T_{\rm eff}$ evolution plots in Figure \ref{fig:inststrip} show where the convective thermal time instability criterion is met for $\ell=1$ and $n=1$ modes.  As is seen, Figure \ref{fig:inststrip} highlights many excellent targets for an observational study of He core WDs, as we comment in the conclusions.


\newpage

\section{Conclusions}
\label{sec:conc}

Our work highlights the $\log g$--$T_{\rm eff}$ parameter space where observable pulsations may be present.  This reveals at least three pulsation candidates, \object[NLTT 11748]{NLTT~11748} \citep{kaw09,ste10}, \object[SDSS J082212.57+275307.4]{SDSS~J0822+2753} \citep{kil09}, and \object[PSR J1012+5307]{PSR~J1012+5307} \citep{van96,cal98}, all of which should be observed for variability on timescales of 200--1000\,s.   Mode detections and a measurement of the mode period spacing would provide key evidence for an He core composition and large radius, as we predict a period spacing of $\approx 90$\,s for $\ell = 1$, whereas in normal ZZ~Cetis, this number is $\approx 50$\,s \citep{kle98,kan05,pec06}.  Our results regarding the region of instability should be confirmed through future nonadiabatic stability analyses, an issue that is beyond the scope of the present paper.

NLTT~11748 is  highlighted in Figure \ref{fig:inststrip} as a candidate for observable pulsations.  Recent observations by \citet{ste10} did not find pulsations down to a 5\,mmag level, however, they did discover it to be the first eclipsing He WD system.  Given its measured $\log g = 6.20$ and $T_{\rm eff} = 8540$\,K, our models predict a total mass of $0.17\,M_{\odot}$ and envelope mass of $3.15 \times10^{-3}\,M_{\odot}$, comparable to that reported by \citet{kaw09}.  Our numerical pulsation analysis reveals that for this object the lowest order $g$-mode ($\ell = 1$ and $n = 1$) has a period of 245\,s but more importantly the mean period spacing is 89\,s for the $\ell = 1$ modes and 51\,s for the $\ell = 2$ modes.  Figure \ref{fig:propdiag} clearly shows these modes to preferentially reside in the core, offering a unique opportunity to probe the core composition of an He WD.

The current candidates were found in surveys that target other phenomena: the Sloan Digital Sky Survey (SDSS; \citealt{kil07}), high velocity stars \citep{kaw06,kaw09}, and companions to pulsars \citep{van96,cal98,bas06}.  However, the survey selection criteria (photometric colors), although incomplete, can favor WDs of higher $T_{\rm eff}$ ($T_{\rm eff} \gtrsim 11,000$\,K in the SDSS for $\log g < 6$, \citealp{kil07}).  Surveys able to select low-gravity WDs down to $T_{\rm eff} = 8,000$\,K will significantly impact the study of pulsating He core WDs.  

\acknowledgments

We thank David Kaplan for useful discussion in our hopes to observe the first He core pulsator.  This work was supported by the National Science Foundation under grants PHY 05-51164 and AST 07-07633.  P.A. is an Alfred P. Sloan Fellow and acknowledges support from the University of Virginia Fund for Excellence in Science and Technology.



\newpage

\begin{thebibliography}

\bibitem[Althaus et al.(2004)]{alt04}
Althaus, L.~G., C{\'o}rsico, A.~H., Gautschy, A., Han, Z., Serenelli, A.~M., \& Panei, J.~A. 2004, MNRAS, 347, 125 

\bibitem[Bassa et al.(2006)]{bas06}
Bassa, C.~G., van Kerkwijk, M.~H., Koester, D., \& Verbunt, F.  2006, A\&A, 456, 295

\bibitem[Bedin et al.(2008a)]{bed08a} Bedin, L.~R., King, I.~R., Anderson, J., Piotto, G., Salaris, M., Cassisi, S., \& Serenelli, A. 2008a, ApJ, 678, 1279 

\bibitem[Bedin et al.(2008b)]{bed08b}
Bedin, L.~R., Salaris, M., Piotto, G., Cassisi, S., Milone, A.~P., Anderson, J., \& King, I.~R. 2008b, ApJ, 679, L29 

\bibitem[Bedin et al.(2005)]{bed05}
Bedin, L.~R., Salaris, M., Piotto, G., King, I.~R., Anderson, J., Cassisi, S., \& Momany, Y. 2005, ApJ, 624, L45

\bibitem[Brassard \& Fontaine(1997)]{bra97}
Brassard, P., \& Fontaine, G. 1997, Astrophys. Space Sci. Libr., 214, 451 

\bibitem[Brassard et al.(1991)]{bra91}
Brassard, P., Fontaine, G., Wesemael, F., Kawaler, S.~D., \& Tassoul, M. 1991, ApJ, 367, 601 

\bibitem[Brickhill(1983)]{bri83}
Brickhill, A.~J.  1983, MNRAS, 204, 537

\bibitem[Brickhill(1991)]{bri91} 
Brickhill, A.~J.  1991, MNRAS, 251, 673

\bibitem[Castanheira \& Kepler(2008)]{cas08}
Castanheira, B.~G., \& Kepler, S.~O. 2008, MNRAS, 385, 430 

\bibitem[Castanheira et al.(2006)]{cas06}
Castanheira, B.~G., et al. 2006, A\&A, 450, 227 

\bibitem[Castanheira et al.(2007)]{cas07}
Castanheira, B.~G., et al. 2007, A\&A, 462, 989 

\bibitem[Callanan et al.(1998)]{cal98}
Callanan, P.~J., Garnavich, P.~M., \& Koester, D.  1998, MNRAS, 298, 207 

\bibitem[Cassisi et al.(2007)]{cass07}
Cassisi, S., Potekhin, A.~Y., Pietrinferni, A., Catelan, M. \& Salaris, M.  2007, ApJ, 661, 1094

\bibitem[Chang \& Bildsten(2003)]{cha03}
Chang, P. \& Bildsten, L.  2003, ApJ, 585, 464

\bibitem[C{\'o}rsico \& Benvenuto(2002)]{cor02}
C{\'o}rsico, A.~H., \& Benvenuto, O.~G. 2002, Ap\&SS, 279, 281

\bibitem[C{\'o}rsico et al.(2004)]{cor04}
C{\'o}rsico, A.~H., Garc{\'{\i}}a-Berro, E., Althaus, L.~G., \& Isern, J. 2004, A\&A, 427, 923 

\bibitem[D'Cruz et al.(1996)]{dcr96}
D'Cruz, N.~L., Dorman, B., Rood, R.~T., \& O'Connell, R.~W. 1996, ApJ, 466, 359

\bibitem[Deloye \& Bildsten(2002)]{del02}
Deloye, C.~J., \& Bildsten, L. 2002, ApJ, 580, 1077 

\bibitem[Dolez \& Vauclair(1981)]{dol81}
Dolez, N., \& Vauclair, G.\ 1981, A\&A, 102, 375 

\bibitem[Dominguez et al.(1999)]{dom99}
Dominguez, I., Chieffi, A., Limongi, M., \& Straniero, O. 1999, ApJ, 524, 226

\bibitem[Driebe et al.(1999)]{dri99}
Driebe, T., Bl{\"o}cker, T., Sch{\"o}nberner, D., \& Herwig, F. 1999, A\&A, 350, 89 

\bibitem[Dziembowski(1977)]{dzi77}
Dziembowski, W.  1977, Acta Astron., 27, 1

\bibitem[Dziembowski \& Koester(1981)]{dzi81}
Dziembowski, W., \& Koester, D.\ 1981, A\&A, 97, 16 

\bibitem[Ferguson et al.(2005)]{fer05}
Ferguson et al.  2005, ApJ, 623, 585

\bibitem[Fontaine \& Brassard(2008)]{fon08}
Fontaine, G., \& Brassard, P.  2008, PASP, 120, 1043 

\bibitem[Fontaine et al.(2003)]{fon03}
Fontaine, G., Brassard, P., \& Charpinet, S. 2003, Ap\&SS, 284, 257


\bibitem[Garc{\'{\i}}a-Berro et al.(2010)]{gar10} 
Garc{\'{\i}}a-Berro, E., et al.\ 2010, Nature, 465, 194 

\bibitem[Gianninas et al.(2007)]{gia07}
Gianninas, A., Bergeron, P., \& Fontaine, G.  2007, in ASP Conf. Ser. 372, 15th European Workshop on White Dwarfs, ed. R. Napiwotzki \& M. R. Burleigh (San Francisco, CA: ASP), 577 

\bibitem[Hansen(2005)]{han05}
Hansen, B.~M.~S.\ 2005, ApJ, 635, 522 


\bibitem[Iben \& Livio(1993)]{ibe93}
Iben, I.~J., \& Livio, M. 1993, PASP, 105, 1373 

\bibitem[Inglesias \& Rogers(1993)]{ing93}
Inglesias, C.~A.  \& Rogers, F.~J.  1993, ApJ, 412, 752

\bibitem[Inglesias \& Rogers(1996)]{ing96}
Inglesias, C.~A.  \& Rogers, F.~J.  1996, ApJ, 464, 943

\bibitem[Kalirai et al.(2007)]{kal07}
Kalirai, J.~S., Bergeron, P., Hansen, B.~M.~S., Kelson, D.~D., Reitzel, D.~B., Rich, R.~M., \& Richer, H.~B. 2007, ApJ, 671, 748 

\bibitem[Kanaan et al.(2005)]{kan05}
Kanaan, A., et al. 2005, A\&A, 432, 219 

\bibitem[Kawaler(1988)]{kaw88}
Kawaler, S.~D. 1988, Apj, 334, 220 

\bibitem[Kawka \& Vennes(2009)]{kaw09}
Kawka, A., \& Vennes, S.  2009, A\&A, 506, L25 

\bibitem[Kawka et al.(2006)]{kaw06}
Kawka, A., Vennes, S., Oswalt, T.~D., Smith, J.~A., \& Silvestri, N.~M.  2006, ApJ, 643, L123


\bibitem[Kilic et al.(2007)]{kil07}
Kilic, M., Allende Prieto, C., Brown, W.~R., \& Koester, D.  2007, ApJ, 660, 1451

\bibitem[Kilic et al.(2010)]{kil09}
Kilic, M., Brown, W.~R., Allende Prieto, C., Kenyon, S.~J., \& Panei, J.~A.\ 2010, ApJ, 716, 122 

\bibitem[Kleinman et al.(1998)]{kle98}
Kleinman, S.~J., et al. 1998, ApJ, 495, 424 

\bibitem[Marsh et al.(1995)]{mar95}
Marsh, T.~R., Dhillon, V.~S., \& Duck, S.~R. 1995, MNRAS, 275, 828 

\bibitem[Mukadam et al.(2004)]{muk04}
Mukadam, A.~S., Winget, D.~E., von Hippel, T., Montgomery, M.~H., Kepler, S.~O., 
\& Costa, A.~F.~M.  2004, ApJ, 612, 1052


\bibitem[Panei et al.(2007)]{pan07}
Panei, J.~A., Althaus, L.~G., Chen, X., \& Han, Z.  2007, MNRAS, 382, 779


\bibitem[Pech et al.(2006)]{pec06}
Pech, D., Vauclair, G., \& Dolez, N. 2006, A\&A, 446, 223 

\bibitem[Pietrinferni et al.(2004)]{pie04}
Pietrinferni, A., Cassisi, S., Salaris, M., \& Castelli, F. 2004, ApJ, 612, 168 

\bibitem[Rogers \& Nayfonov(2002)]{rog02}
Rogers, F.~J. \& Nayfonov, A.  2002, ApJ, 576, 1064

\bibitem[Saumon et al.(1995)]{sau95}
Saumon, D., Chabrier, G. \& Van Horn, H.~M.  1995, ApJS, 99, 713

\bibitem[Serenelli et al.(2002)]{ser02}
Serenelli, A.~M., Althaus, L.~G., Rohrmann, R.~D., \& Benvenuto, O.~G. 2002, MNRAS, 337, 1091 

\bibitem[Steinfadt et al.(2008)]{ste08}
Steinfadt, J.~D.~R., Bildsten, L., Ofek, E.~O., \& Kulkarni, S.~R. 2008, PASP, 120, 1103 

\bibitem[Steinfadt et al.(2010)]{ste10}
Steinfadt, J.~D.~R., Kaplan, D.~L., Shporer, A., Bildsten, L., \& Howell, S.~B.\ 2010, ApJ, 716, L146 

\bibitem[Timmes(1999)]{tim99}
Timmes, F.~X.  1999, ApJS, 124, 241

\bibitem[Timmes \& Swesty(2000)]{tim00}
Timmes, F.~X. \& Swesty, F.~D.  2000, ApJS, 126, 501

\bibitem[Unno et al.(1989)]{unn89}
Unno, W., Osaki, Y., Ando, H., Saio, H., \& Shibahashi, H. 1989, \textit{Nonradial Oscillations of Stars}, (2nd ed.; Tokyo: Univ. of Tokyo Press)  

\bibitem[van Kerkwijk et al.(1996)]{van96}
van Kerkwijk, M.~H., Bergeron, P., \& Kulkarni, S.~R.  1996, ApJ, 467, L89 

\bibitem[Warner \& Robinson(1972)]{war72}
Warner, B., \& Robinson, E.~L. 1972, Nature. Nat. Sci., 239, 2 

\bibitem[Wesemael et al.(1991)]{wes91}
Wesemael, F., Bergeron, P., Fontaine, G., \& Lamontagne, R. 1991, in NATO ASIC Proc.~336: White Dwarfs, 7th European Workshop, ed. G. Vauclair \& E. Sion (Dordrecht: Kluwer), 159

\bibitem[Winget \& Kepler(2008)]{win08}
Winget, D.~E., \& Kepler, S.~O. 2008, ARA\&A, 46, 157

\bibitem[Winget et al.(1982)]{win82}
Winget, D.~E., van Horn, H.~M., Tassoul, M., Fontaine, G., Hansen, C.~J., \& Carroll, B.~W.\ 1982, ApJ, 252, L65

\bibitem[Wu \& Goldreich(1999)]{wu99}
Wu, Y., \& Goldreich, P.  1999, ApJ, 519, 783


\end{thebibliography}
\end{document}